\documentclass{JHEP3}
\usepackage{epsfig}
\usepackage{amsbsy}
\usepackage{varioref}
\usepackage{pifont}

\newcommand{\hs}{\hspace*{0.5cm}}

\newcommand{\be}{\begin{equation}}
\newcommand{\ee}{\end{equation}}
\newcommand{\bea}{\begin{eqnarray}}
\newcommand{\eea}{\end{eqnarray}}
\newcommand{\bary}{\begin{array}}
\newcommand{\eary}{\end{array}}
\newcommand{\bit}{\begin{itemize}}
\newcommand{\eit}{\end{itemize}}
\newcommand{\ben}{\begin{enumerate}}
\newcommand{\een}{\end{enumerate}}
\newcommand{\crn}{\nonumber \\}
\newcommand{\nn}{\nonumber}

\newcommand{\al}{\alpha}
\newcommand{\la}{\lambda}
\newcommand{\bet}{\beta}
\newcommand{\ga}{\gamma}

\newcommand{\om}{\omega}

\newcommand{\fr}{\frac}

\newcommand{\bc}{\begin{center}}
\newcommand{\ec}{\end{center}}

\newcommand{\ep}{\epsilon}

\newcommand{\La}{\Lambda}

\newcommand{\si}{\sigma}

\newcommand{\Om}{\Omega}

\def\sla#1{\ifmmode%
\setbox0=\hbox{$#1$}%
\setbox1=\hbox to\wd0{\hss$/$\hss}\else%
\setbox0=\hbox{#1}%
\setbox1=\hbox to\wd0{\hss/\hss}\fi%
#1\hskip-\wd0\box1 }
\title{Photon-radion conversion cross-sections in
external electromagnetic field }
\author{ P. V.  Dong$^a$, H. N. Long$^a$,  D.  V. Soa$^b$ and N. H.Thao$^a$ \\
  $^a$ Institute of  Physics, VAST, 10 Dao Tan, Ba Dinh, Hanoi,
  Vietnam\\  $^b$ Department of Physics, Hanoi University of
Education, Hanoi, Vietnam\\
 E-mail: \email{pvdong@iop.vast.ac.vn},
 \email{hnlong@iop.vast.ac.vn},
 \email{dvsoa@assoc.iop.vast.ac.vn},
\email{nhthao@grad.iop.vast.ac.vn} }

\abstract{An attempt is made to present some experimental
predictions of the Randall-Sundrum model, where compactification
radius of the extra dimension is stabilized by the radion, which
is a scalar field lighter than the graviton Kaluza-Klein states.
We calculate  the conversion cross-sections of the photons into
the radions in the  external electromagnetic fields, namely, in
the static fields and in a periodic field of the wave guide.
Numerical evaluations of the total cross-sections are also given.
Our result shows that the conversion cross-section in the static
electric field is quite small. But, in the static magnetic and
periodic fields, the radion productions are much enhanced.}


\begin{document}


\maketitle

\section{Introduction}

There has been a lot attention devoted to the models of physics above
the weak scale utilizing extradimensions in solving the hierarchy
problem. Firstly, Arkani-Hamed, Dimopoulos and
Dvali~\cite{add} have suggested that the fundamental scale of quantum
gravity could be dramatically much lower than the Planck scale provided
the standard model (SM) fields propagate on a 3-dimensional brane
and the gravity propagates in extraspace dimensions. The smallness of
Newton's constant can then be explained by the large size of the
volume of compactification. An alternative approach as proposed by Randall and Sundrum (RS)
\cite{rs} can also solve the hierarchy problem by localizing all the SM
particles on the visible brane or TeV brane of a non-factorizable
geometry like a slice of five-dimensional (5D) anti de-Sitter
space with curvature $\kappa$. The fifth dimension of
this space is $S^1/Z_2$ orbifold of size $r$ labeled by a
coordinate $y \in [0,1/2 ]$, such that the points $(x^\mu, y)$ and
$(x^\mu, -y)$ are identified. The four-dimensional (4D) massless
graviton field is localized on another brane away from our
brane. Since the fundamental
gravity scales as motivated by the hierarchy problem were not much bigger than a TeV, these scenarios
would have distinctive signatures in collider experiments.

In the RS model the compactification radius is in the
order of Planck length and interestingly is a dynamical object. It
is connected to the vacuum expectation value (VEV) of the dilaton
field arising due to the compactification of full 5D theory to 4D.
The radion field is an exponential function of the dilaton field scaled
by proper factors \cite{csaba1}. Much research has been done on
understanding a possible mechanism for radius stabilization and the
phenomenology of the radion field in the
model~\cite{rradion}. The motivation for studying the radion is
twofold. Firstly, the radion may turn out to be the lightest new
particle in the RS-type setup, possibly accessible at the LHC. In
addition, the phenomenological similarity and potential mixing of
the radion and Higgs boson warrant detailed study in order to
facilitate distinction between the radion and Higgs signals at
colliders.

One of the intriguing features of the RS models is that the
neutral scalar bosons can have interactions with the photon. This
feature has been recently investigated in a series of
works~\cite{mada,rhiggs1,chi,rhiggs,csaki,phonvh,gon,hoom}. The couplings
of two photons with neutral scalar bosons provide interesting
possibility to search the latter in the external electromagnetic
(EM) fields. The light particles with a two photon interaction may
be created by a photon entering the EM  field, an effect firstly
discussed by Primakoff ~\cite{pri}. This effect is the basis of
Sikivie's methods for the detection of axions and other light
particles in a resonant cavity~\cite{sik}. The pseudoscalar--photon mixing
phenomenon in background magnetic field has been
analyzed in detail in the literature~\cite{mir}. This phenomenon
has also been used to impose stringent limits on the pseudoscalar--photon coupling~\cite{and}.  In the previous works, we have
considered the creation of gravitons and
dilations~\cite{lhmpla,lst} and axions~\cite{plb95} in the
external EM field.

The aim of this work is to study the phenomenology of radion in
the RS model and the conversion possibility of photon into radion
in the external EM fields. In contradiction with the previous
experiments of gravitons, dilations and axions, the radion in the
RS model may be much heavier. Therefore, we should take into
account the provided source of high energy photons. One way to
achieve the high energy photons is to use the laser backscattering
technique~\cite{gin}. Then the light radion (in order of a few
GeV) of the RS model can be produced.

The organization of this paper is follows. In Sec. \ref{rsmodel}
we give a review of the RS model. In Sec. \ref{rphph} we present
the coupling of radion to photons. Section
\ref{rprod} is devoted to photon-radion conversions in the
external EM fields. Namely, we first account for the production of
radions in an 100 KV/m external electric field of flat condenser
and the case in a 9 Tesla external magnetic field of solenoid in
CAST experiments \cite{cast}. We next consider the conversion in a
periodic EM field of wave guide. Finally, we summarize our results
and make conclusions in the last section--Sec. \ref{culc}.

\section{\label{rsmodel}A review of RS  model}
The RS model is based on a 5D spacetime with non-factorizable
geometry \cite{rs}. The single extradimension is compactified on a
$S^1/Z_2$ orbifold of which two fixed points accommodate two
three-branes (4D hyper-surfaces), the Planck brane at $y= 0$ and
TeV brane at $y = 1/2$. The ordinary 4D Poincare invariance is
shown to be maintained by the following classical solution to the
Einstein equation: \be ds^2 = e^{-2\si(y)}\eta_{\mu \nu} dx^\mu
dx^\nu - b_0^2 d y^2, \hs \si(y) = m_0 b_0 |y|, \label{rsdt1} \ee
where $x^\mu$ $(\mu=0,1,2,3)$ denote the coordinates on the 4D
hyper-surfaces of constant $y$ with metric $\eta_{\mu \nu}=
\mathrm{diag}(1,-1,-1,-1) $. The $m_0$ and $b_0$ are the
fundamental mass parameter and compactification radius,
respectively.

 Gravitational fluctuations about the RS metric,
\be \eta_{\mu \nu} \rightarrow g_{\mu \nu} = \eta_{\mu \nu} + \ep
h_{\mu \nu}(x,y), \hs b_0 \rightarrow b_0 + b(x),\hs \ \ee yield
two kinds of new phenomenological ingredients on the TeV brane:
the KK graviton modes $h^{(n)}_{\mu \nu}(x)$ and the canonically
normalized radion field $\phi_0(x)$, respectively defined as \be
h_{\mu \nu}(x,y) =\sum_{n=0}^\infty h^{(n)}_{\mu \nu}(x)
\fr{\chi^{(n)}(y)}{\sqrt{b_0}},\hs
\phi_0(x)=\sqrt{6}M_{\mathrm{Pl}}\Om_b(x), \ee
 where $\Om_b(x) \equiv e^{-m_0[b_0 +b(x)]/2}$.
 The 5D Planck mass $M_5\ (\ep^2 = 16 \pi G_5 = 1/M^3_{5})$ is
 related to its 4D one ($M_{\mathrm{Pl}} \equiv 1/\sqrt{8 \pi G_\mathrm{N}}$) by
\be \fr{M^2_{\mathrm{Pl}}}{2} = \fr{1 - \Om^2_0 }{\ep^2 m_0}. \ee
Here $\Om_0 \equiv  e^{-m_0 b_0/2}$ is known  as the warp factor.
Because our TeV brane is arranged to be at $y = 1/2$, a
canonically normalized scalar field has the mass multiplied by the
warp factor, i.e, $m_{\mathrm{phys}}=\Om_0 m_0 $.
 Since the moderate value of  $m_0b_0/2 \simeq 35$ can generate TeV
 scale physical mass, the gauge hierarchy problem is explained.

The 4D effective Lagrangian is then \be \mathcal{L} =
-\fr{\phi_0}{\La_\phi}T^\mu_\mu -\fr{1}{\hat{\La }_W} T^{\mu
\nu}(x)\sum_{n=1}^\infty h^{(n)}_{\mu \nu}(x), \ee  where
$\La_\phi\equiv\sqrt{6}M_{\mathrm{Pl}}\Om_0$ is the VEV of the
radion field, and $\hat{\La}_W\equiv\sqrt{2}M_{\mathrm{Pl}}\Om_0$.
The $T^{\mu \nu}$ is the energy-momentum tensor of the TeV brane
localized SM fields. The $T^\mu_\mu$ is the trace of the
energy-momentum tensor, which is given at the tree level
as~\cite{sae,csabm} \be T^\mu_\mu = \sum_{f} m_f \bar{f}f - 2
m^2_W W^{+}_{\mu}W^{-\mu} - m^2_Z Z_{\mu}Z^{\mu}+(2m^2_{h_0} h_0^2
-\partial_{\mu}h_0 \partial^{\mu}h_0) + \cdots \ee

The gravity-scalar  mixing arises at the  TeV-brane
by~\cite{rhiggs1,chi,rhiggs} \be S_\xi = -\xi \int d^4 x
\sqrt{-g_{\mathrm{vis}}} R(g_{\mathrm{vis}}) \hat{H}^\dagger
\hat{H}, \ee where $R(g_{\mathrm{vis}})$ is the Ricci scalar for
the induced metric on the visible brane or TeV brane,
$g_{\mathrm{vis}}^{\mu \nu} = \Om^2_b(x)(\eta^{\mu \nu} + \ep
h^{\mu \nu})$. $\hat{H}$ is the Higgs field before re-scaling,
i.e., $ H_0 = \Om_0 \hat{H}$. The parameter $\xi$ denotes the size
of the mixing term. With $\xi \neq 0$, there is neither a pure
Higgs boson nor pure radion mass eigenstate. This $\xi$ term mixes
the $h_0$ and $\phi_0$ fields into the mass eigenstates $h$ and
$\phi$ as given by~\cite{chi,rhiggs} \bea
\left(%
\begin{array}{c}
  h_0 \\
  \phi_0\\
\end{array}%
\right) & = &
\left(%
\begin{array}{cc}
 1 & -6\xi \ga/Z \\
 0 & 1/Z \\
\end{array}%
\right)
\left(%
\begin{array}{cc}
\cos \theta & \sin \theta \\
 -\sin \theta &  \cos \theta \\
\end{array}%
\right) \left(%
\begin{array}{c}
 h \\
 \phi\\
\end{array}%
\right)=
\left(%
\begin{array}{cc}
   d &  c \\
 b &  a \\
\end{array}%
\right) \left(%
\begin{array}{c}
   h \\
  \phi\\
\end{array}%
\right), \label{rsd5a}
 \eea where
 \bea
\ga &\equiv& v_0/\La_\phi,\hs Z^2  \equiv  1 - 6 \xi \ga^2 (1+6
\xi) = \bet - 36 \xi^2 \ga^2,\hs
 \bet \equiv 1 - 6 \xi \ga^2,\crn
 a&\equiv&\fr{\cos \theta}{Z}, \hs
b\equiv -\fr{\sin \theta}{Z},\hs c\equiv \sin \theta - \fr{6\xi
\ga}{Z}\cos \theta,\hs d \equiv\cos \theta +
 \fr{6\xi \ga}{Z}\sin \theta. \label{rsd2}\eea The mixing angle $\theta$ is defined by
\bea \tan 2 \theta &=& 12 \ga \xi Z \fr{m^2_{h_0}}{m^2_{h_0}(Z^2
-36 \xi^2 \ga^2)-m^2_{\phi_0}}. \label{rsd3} \eea The new fields
$h$ and $\phi$ are mass eigenstates with masses \be m^2_{h,\phi} =
\fr{1}{2 Z^2}\left[m^2_{\phi_0} + \bet m^2_{h_0} \pm
\sqrt{(m^2_{\phi_0}+ \bet m^2_{h_0} )^2 - 4 Z^2
m^2_{\phi_0}m^2_{h_0} }\right]. \label{rsd4} \ee

The mixing between the states enable decays of the heavier
eigenstate into the lighter eigenstates if kinematically
 allowed. Overall, the production cross-sections, widths and relative branching fractions can all be affected
 significantly by the value of the mixing parameter $\xi$~\cite{rhiggs1,rhiggs, csabm, cheung}.
 There are also two algebraic constraints on the value of $\xi$. One comes from the requirement that
the roots of the inverse functions of Eq. (\ref{rsd4}) are
definitely  positive. Suggesting that the Higgs boson is heavier,
we get \bea \frac{m^2_{h}}{m^2_{\phi}} > 1+
\frac{2\bet}{Z^2}\left(1- \frac{Z^2}{\bet}\right) +
\frac{2\bet}{Z^2} \left[1-
\frac{Z^2}{\bet}\right]^{1/2}\label{rsd41}.\eea The other one is
from the fact that the
 $Z^2$ is the coefficient of the radion kinetic term after undoing the
 kinetic mixing. It is therefore required to be positive ($Z^2 > 0$) in order to keep
 the radion kinetic term  definitely  positive, i.e. \bea
-\frac{1}{12}\left(1+\sqrt{1+\frac{4}{\ga^2}}\right)
 < \xi < \frac{1}{12}\left(\sqrt{1+\frac{4}{\ga^2}}-1\right) \label{rsd42} . \eea

We now discuss the previous estimations on the radion mass  and
some model parameters. All phenomenological signatures of the RS
model including the radion - Higgs mixing are specified by five
parameters \be \La_\phi , \ \fr{m_0}{M_{\mathrm{Pl}}}, \ m_h , \
m_\phi , \ \xi. \ee For  the reliability of the  RS  solution, the
ratio $\ \fr{m_0}{M_{\mathrm{Pl}}}$ is usually taken around
$0.01\leq\ \fr{m_0}{M_{\mathrm{Pl}}}\leq 0.1$ to avoid too large
 bulk curvature ~\cite{davou}. Therefore, we
consider the case of  $\Lambda_\phi = 5\ \textrm{TeV}$ and $\
\fr{m_0}{M_{\mathrm{Pl}}}= 0.1$, where the effect of radion on the
oblique parameters is small~\cite{cskim}. The radion mass in the
RS model is expected to be light as by one of the simplest
stabilization mechanisms predicted  in range of $10 - 100 \
\textrm{GeV}$~\cite{rradion}. We choose $\xi = 0,
\pm1/6$, which are in agrement with those in Ref.~\cite{rhiggs1} with $\xi
\ga\ll 1, Z^2\approx 1$.

The recent results in
Refs.~\cite{csabm,anz} have shown that the radion can be naturally
stabilized with a smaller mass, for example, in order of $10^{-2}\
\textrm{GeV}$. Perhaps a much lower mass can also be accommodated
with little fine-tuning, which is necessary for the conversion processes
considered below to be more relevant, but in general the radion
is not naturally such small. There is nothing wrong with finding
that the experiments do not yet probe the theoretically expected
parameter space of the certain models, but that it comes close,
and therefore it may be worth looking for the radion in this way.
In this work, we will show the cross-section expressions for the
general cases with arbitrary radion mass, but the numerical
treatments only take the GeV radions into account where the effect
of Higgs-radion mixing may become important. (Let us recall that in the large extradimensions, the radion is
typically very light, with mass between $10^{-3}\ \textrm{eV}$ and
$\textrm{MeV}$.)

\section{\label{rphph}Radion coupling to photons}

For the massless gauge bosons such as photon and gluon, there are
no large couplings to the radion because there are no
brane-localized mass terms. However, the potentially large
contributions to these couplings may come from the loop effects of
the gauge bosons, the higgs field and the top quark as well as the
localized trace anomalies (there are also the bulk contributions
if the massless gauge bosons are set off-brane, but this does not
change the main results of the paper)
\cite{rhiggs1,chi,csaki,kin}.

Referring the reader for details of the radion-photon coupling to
Ref.~\cite{rhiggs1,chi}, we lay out the necessary radion-photon
coupling \bea
 \mathcal{L}_{\ga \ga \phi}& = & \fr 1 2  c_{\phi \ga \ga} \phi F_{\mu \nu}F^{\mu \nu}, \label{rsd6}
\eea with \be c_{\phi \ga \ga} =  -\fr{\al}{4 \pi
\La_\phi}\left\{a(b_2+ b_Y) -  a_{12}[F_1(\tau_W) +
 4/3 F_{1/2}(\tau_t)]\right\},\ee
 where $b_2= 19/6, b_Y = -41/6$ are the $\mathrm{SU}(2)_L\otimes
 \mathrm{U}(1)_Y$ $\beta$-function coefficients in the SM, and
 $a_{12}= a + c/\gamma$, $\tau_{t}=4m^2_{t}/q^2$ and $\tau_{W}=4m^2_{W}/q^2$.

The form factors $F_{1/2}(\tau_t)$ and $F_1(\tau_W)$ are given by
\bea F_{1/2}(\tau) & =& -2 \tau[1+(1-\tau) f(\tau)],\crn F_1(\tau)
&=& 2 + 3\tau+3\tau(2-\tau) f(\tau), \label{rsd7} \eea with \bea
f(\tau) = \left\{
\begin{array}{l}
\arcsin^2(1/\sqrt{\tau}),\hs \hs \hs \hs \hs  \tau \geq 1 ,
\\
-\fr 1 4 \left[\ln\left( \fr{1+\sqrt{1-\tau}}{1-\sqrt{1-\tau}}
\right) - i \pi \right]^2  \hs \hs \! \tau < 1  .
\end{array}
\right. \eea The important property of $ F_{1/2}(\tau)$ is that,
for $\tau > 1$, it very quickly saturates to $-4/3$, and to $0$
for $\tau < 1$.  $F_1(\tau)$ saturates quickly to $7$ for $\tau >
1$, and to $0$ for $\tau < 1$ \cite{csaki}.

Now let us turn to our main interest, i.e., the photon conversions
into radions in the external EM  fields.

\section{\label{rprod}Photon-to-radion conversions}

The photon regeneration experiment, using RF photons, was
described in Ref.~\cite{hoog}. That experiment consists of two
cavities which are placed a small distance apart. A more or less
homogeneous magnetic field exists in both the cavities. The first, or
emitting  cavity,  is excited by incoming RF radiation. Depending
on the radion-photon coupling constant, a certain amount of RF
energy will be deposed in the second,  or receiving cavity. Using
Feynman diagram technique we have considered the conversion of the
photon into axions in external EM  field~\cite{plb95}. In this
paper, in the framework of the RS model we apply this method to
reconsider the conversion of photon into the radion.

Let us consider the conversion of the photon $\ga$ with momentum
$q$ into radion  $\phi$ with momentum $p$ in an external EM field.
Using the Feynman rules we get the following expression for the
matrix element \bea < p  | M_\phi | q > = \frac{c_{\phi \ga
\ga}}{(2\pi)^2 \sqrt{p_0 q_0}}\varepsilon^\mu({\bf q},\la) q^\nu
\int_V e^{i{\bf k r} }F^{\mathrm{class}}_{\nu \mu} d {\bf r},
\label{rsd31} \eea where $\textbf{k} \equiv \textbf{p} -
\textbf{q}$ and $\varepsilon^\mu({\bf q},\la)$ represents the
polarization vector of the photon. Expression (\ref{rsd31}) is
valid for an arbitrary external EM field. In the following we
shall use it for the cases, namely in the electric field of a flat
condenser, in the static magnetic field of a solenoid and in a
wave guide with the $\mathrm{TE}_{10}$ mode. Here we use the
following notations: $q \equiv |\mathbf{q}|$, $p\equiv
|\mathbf{p}| = (q^2 -m_\phi^2)^{1/2}$ and $\theta$ is the angle
between $\mathbf{p}$ and $\mathbf{q}$.

\subsection{Conversion in electric field}

Let us take the EM field as a homogeneous electric field of a flat
condenser of size $l_x \times l_y \times l_z$. We  shall use the
coordinate system with the $x$ axis parallel to the direction of
the field, i.e., $F^{01}= - F^{10} = E$. Then the matrix element
is given by \bea < p | M_\phi | q > = \frac{c_{\phi \ga
\ga}}{(2\pi)^2 \sqrt{p_0 q_0}}\varepsilon^1({\bf q},\la) q^0
F_e(\textbf{k}), \label{rsd32} \eea where \bea F_e(\textbf{k})=
\int_{V} e^{i{\bf k r}}E(\mathbf{r}) d\mathbf{r}\label{rsd401}
\eea For a homogeneous electric field of intensity $E$ we have \be
F_e(\textbf{k}) = 8 E
\sin(l_xk_x/2)\sin(l_yk_y/2)\sin(l_zk_z/2)(k_x k_y k_z)^{-1}.
\label{rsd34} \ee Squaring the matrix element (\ref{rsd32}) we
obtain
\be\frac{d\sigma^e(\gamma\rightarrow\phi)}{d\Omega}=\frac{8c^2_{\phi
\ga \ga}E^2 q^2}{\pi^2}\left[\frac{\sin(\frac{1}{2}l_xk_x)
\sin(\frac{1}{2}l_yk_y)\sin(\frac{1}{2}l_zk_z)}{k_xk_yk_z}\right]^2\left(1-\frac{q^2_x}{q^2}\right).
\label{rsd35} \ee
 From Eq.(\ref{rsd35}) we see that if the photon moves in the
direction of the electric field, i.e., $q^\mu =(q,q,0,0)$, the
differential cross-section vanishes.

We  shall explore the following  case: The momentum of photon is
parallel to the $z$ axis, i.e., $q^\mu =(q,0,0,q)$. In the
spherical  coordinates we then have \be p_x= p\sin\theta \cos
\varphi, \hs p_y= p\sin\theta \sin \varphi,\hs  p_z= p\cos\theta,
\label{rsd36} \ee where $\varphi$ is the angle between the $x$
axis and the projection of $\textbf{p}$ on the $xy$ plane.
Substitution of  Eq.(\ref{rsd36}) into Eq.(\ref{rsd35}) yields
\bea \fr{d\si^e(\ga \rightarrow \phi )}{d \Omega} & = & \fr{8
c_{\phi \ga \ga}^2 E^2 q^2 }{ \pi^2} \left[
 \sin \fr{l_x p \sin \theta \cos \varphi}{2} \sin \fr{l_yp \sin \theta \sin
 \varphi}{2}
 \sin \fr{l_z(q-p \cos \theta)}{2}\right]^2\crn
&\times& [p^2(q-p \cos \theta)\sin^2 \theta \cos \varphi \sin
\varphi]^{-2}. \label{rsd37} \eea

Let us evaluate the differential cross-section in the following
several limits. When the scattering angle $\theta$ is very small,
i.e., $\sin\theta\approx\theta $,  we get then \bea \fr{d\si^e(\ga
\rightarrow \phi )}{d \Omega} & = & \fr{c_{\phi \ga \ga}^2 E^2
l_x^2 l_y^2 }{ 2\pi^2
\left(1-\sqrt{1-\fr{m_\phi^2}{q^2}}\right)^2}
 \sin^2\left[
 \fr{l_zq}{2}\left(1-\sqrt{1-\fr{m_\phi^2}{q^2}}\right)\right].
\label{rsd38} \eea In the limit $\theta \rightarrow\fr{\pi}{2}$
and $\varphi \rightarrow 0$, the differential cross-section
(\ref{rsd37}) becomes \bea \fr{d\si^e(\ga \rightarrow \phi )}{d
\Omega} & = & \fr{2 c_{\phi \ga \ga}^2 E^2 l_y^2 }{ \pi^2
(q^2-m_\phi^2)} \sin^2 \fr{l_xq\sqrt{1-\frac{m_\phi^2}{q^2}}}{2}
 \sin^2\fr{l_zq}{2}.\label{rsd39} \eea
When $\theta \rightarrow\fr{\pi}{2}$ and $\varphi \rightarrow
\fr{\pi}{2}$, Eq.(\ref{rsd37}) gives \bea \fr{d\si^e(\ga
\rightarrow \phi )}{d \Omega} & = & \fr{2 c_{\phi \ga \ga}^2 E^2
l_x^2 }{ \pi^2 (q^2-m_\phi^2)} \sin^2
\fr{l_yq\sqrt{1-\frac{m_\phi^2}{q^2}}}{2}
 \sin^2\fr{l_zq}{2},\label{rsd40} \eea which is
 similar to Eq.(\ref{rsd39}) with $l_x\leftrightarrows l_y$.

From  Eqs.(\ref{rsd38},\ref{rsd39},\ref{rsd40}) we see that the
differential cross-sections in the direction of the condenser
depends quadratically on the intensity $E$, the sizes of
condenser, and the photon momentum $q$. Since the external EM
field is classical we can therefore increase the scattering
probability as much as possible by increasing the intensity of the
field and/or the condenser volume containing the field.

Now we are interested in the total cross-section $\sigma^e(q)=\int
d \Omega (d \sigma^e /d\Omega)$. Since the integrand as given by
the general formula (\ref{rsd37}) does not simultaneously vanish
over the integrated domain, the total cross-section is always
different from (i.e. larger than) zero. Because the integrand as
well as the total cross-section which depend on provided photon
high momentum $q$ (at least larger than the radion mass) are very
rapidly oscillated with $q$ (an evaluation for $l_{x,y,z}q$ from
the values given below implies this), the numerical treatments are
actually issued by the following problems: \ben\item A plot
constructed from a finite number of points where the neighboring
ones (point next to point) are connected by a line would not have
an obvious variation rule. In principle it is an arbitrary line,
not reflecting the realistic variation of the cross-section or the
spectrum.
\item An average cross-section,
$\bar{\sigma}^e(\bar{q})=\fr{1}{q_2-q_1}
\int^{q_2}_{q_1}dq\sigma^e(q)$, also changes arbitrarily due to
the current numerical methods calculating the integral with a
finite number of divisional points in the integrated domain. \een
To overcome these difficulties, we will plot a large spectrum of
point $(q,\sigma^e(q))$ corresponding to a large number of values
of $q$ in the interested domain. The orientation of the spectrum
will reflect the correct variation of the cross-section. Let us
note that the resulting cross-section will be almost independent
of the radion mass values if $m^2_\phi/q^2 \ll 1$.

\begin{figure}[h]\bc
\includegraphics[width=12cm,height=7cm]{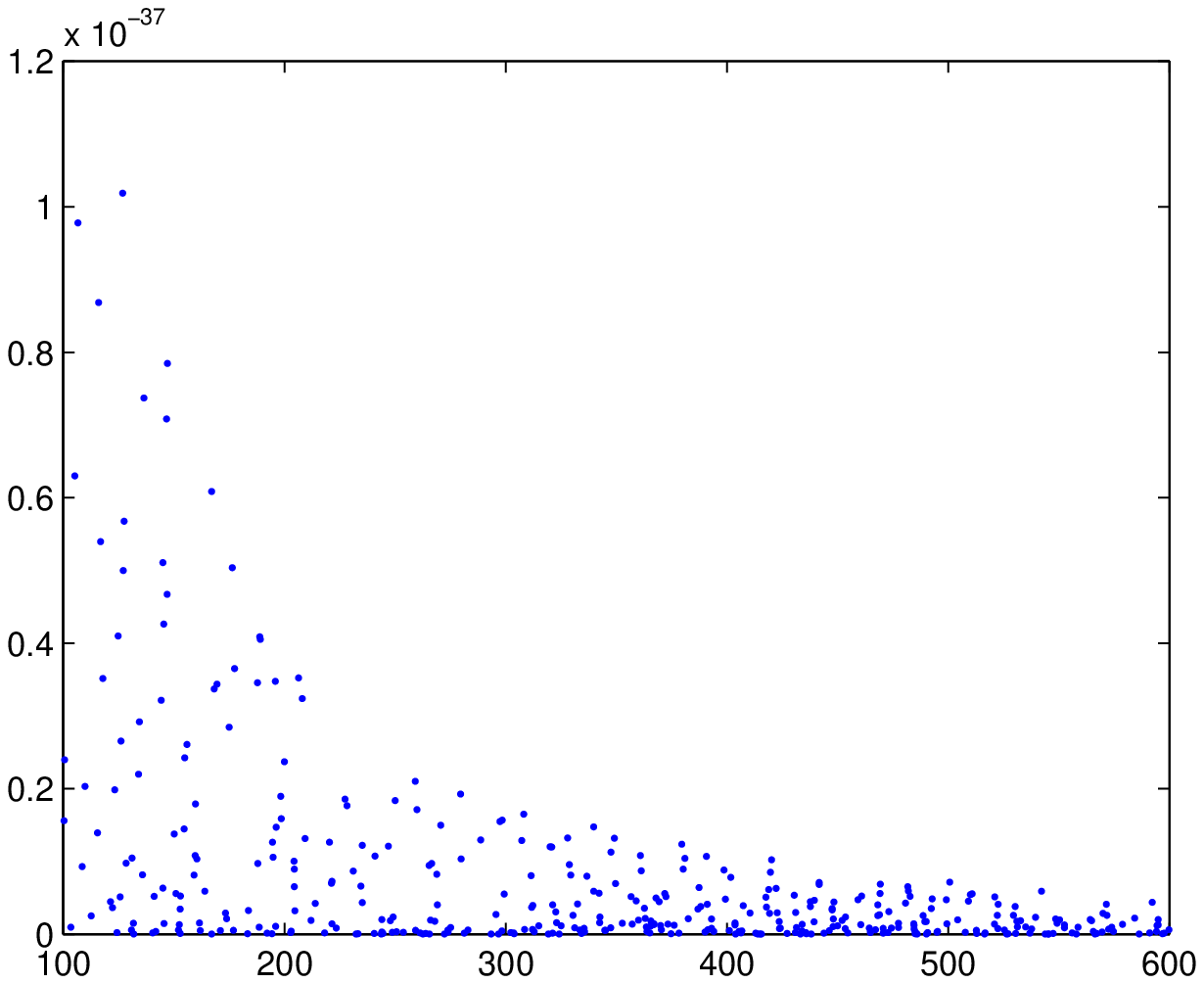}
\includegraphics[width=12cm,height=7cm]{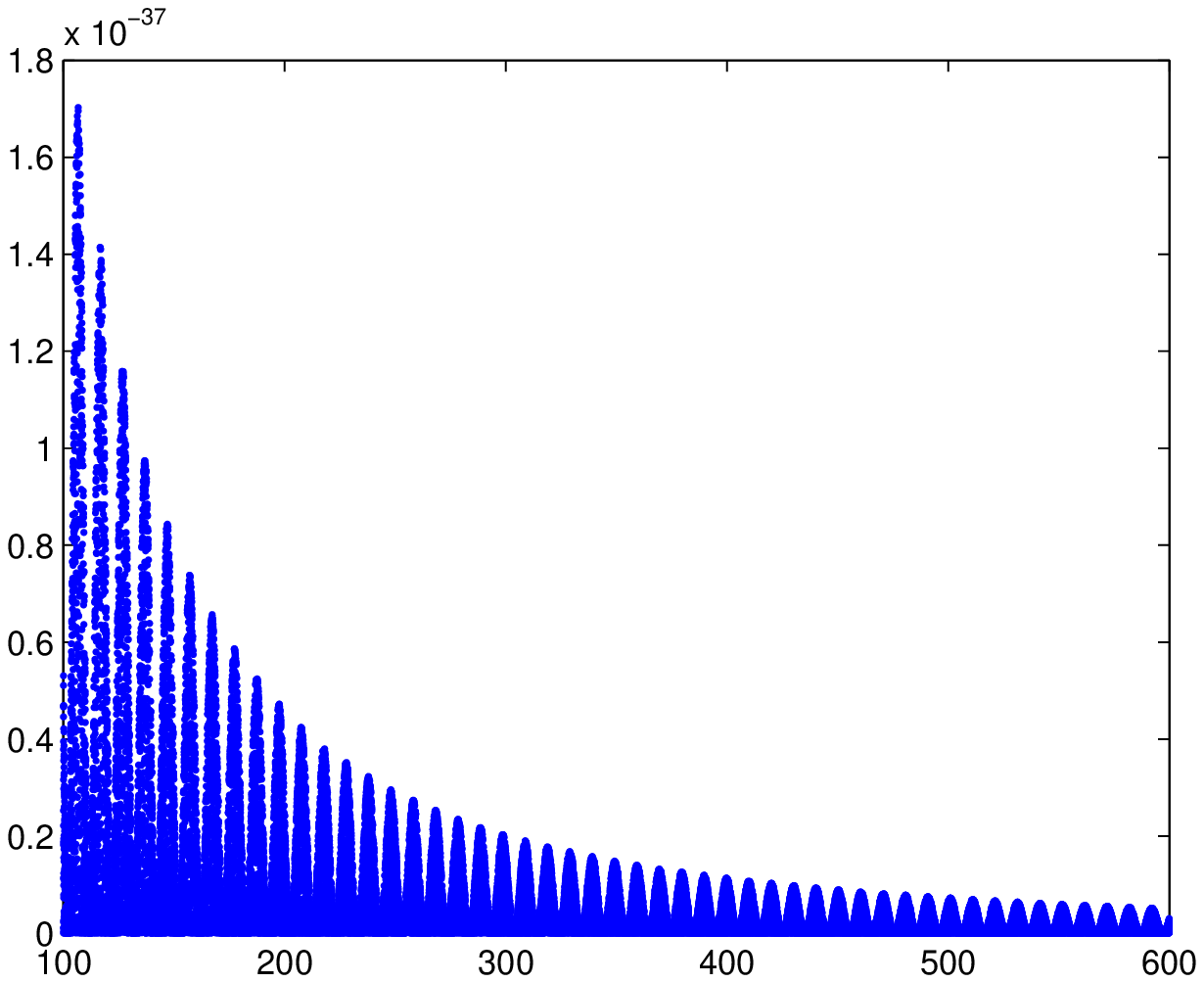}
\caption{\label{tab1}The total cross-section (cm$^2$) for
conversion in electric field as a function of provided momentum
$q=100-600$ GeV. The upper plot is depicted as 400 points, and the
lower one is for 30000 points.} \ec
\end{figure}

In practice, to evaluate the total cross-section
 for Eq.(\ref{rsd37}),  the parameters are chosen as follows:
 $\La_\phi = 5\ \mathrm{TeV}, \xi = 0, \pm1/6, \al = 1/128$, $l_x = l_y = l_z=1\ \mathrm{m}=5.07
 \times 10^6\ \mathrm{eV}^{-1}$,
 $E =100\ \mathrm{KV/m}=6.517\times 10^{-2}\ \mathrm{eV}^2$~\cite{plb95}, and the radion mass
can be taken in the limit  $m_\phi=10 \ \mathrm{
GeV}$~\cite{csabm}. The total cross-section on the selected range
of momenta $q$ for the radion production are given in
Figure~\ref{tab1}. Here the different values $\xi=0,\pm 1/6$
approximately yield the same contribution to the cross-section. As
demonstrated in the two plots, when the number of points is
increased the resonances become shapely. We can see from Figure
\ref{tab1} that the cross-section is quite small to be measurable
because of the current experimental limits, even though the
resonances presented in this case.

In addition, it follows from Figure \ref{tab1} that the changing
of the cross-section in the domain $100\ \textrm{GeV} < q < q_t \
\textrm{GeV} $ is much faster than the remaining domain $q
>  q_t$ GeV, where the ``transition point'' $q_t$ is around on $200-300$ GeV.
In the second one, the cross-section approaches constant
when $q$ is large. The reason for this is that the total
scattering cross-section is dominated by forward scattering at
large $q$. This can be seen in the three subcases of Eqs.
(\ref{rsd38},\ref{rsd39},\ref{rsd40}). Up to rapidly oscillating
(and bounded) terms, the latter two behave like $1/q^2$, whereas
the first one is essentially $q$-independent in this limit
(averaging over oscillations). This can be seen directly also from
the term in square brackets in Eq. (\ref{rsd35}). Due to the
chosen geometry, $k_x$ and $k_y$ are always large ($\sim q$), but
$k_z$ can be small for forward scattering, and thus $\sin k_z/k_z
\rightarrow 1$.

The slope of the cross-section with the large $q$ can be
approximated as $\sigma'(q)\simeq 4.142\times10^{-61}\
\mathrm{cm}^3$, which may be seen from the Figure \ref{tab1}. This
value justifies again that the cross-section tends to the
constant. The transition point $q_t$ can be naively evaluated as
intersection of the asymptote of the cross-section in the large
$q$ domain and the horizontal axis, that is $q_{t}\simeq q-
\fr{\sigma(q)}{\sigma'(q)}\simeq 230\ \mathrm{GeV}$ (valid for
large $q$). It is noticed that this parameter is the derivative
one, not concerning as any characteristic scales of the model. Its
value depends only on a choice of the parameters such as the
radion mass, the size of condenser, the field strength, and so on.

Let us remark that when the momentum of photon is perpendicular to
the electric field $E$ we have then the most optimal condition for
the experiments.

\subsection{Conversion in magnetic field}

Next, we consider the conversion of photon into radion in a
homogeneous magnetic field of the solenoid with  radius \emph{R}
and a length  $l$.  Without loss of generality we suppose that the
direction of the magnetic field is parallel to the z-axis, i.e.,
$F^{12}=-F^{21}=B$. The matrix element is given then \bea <p\mid
M\mid q>=\frac{c_{\phi \ga \ga}}{(2\pi)^2 \sqrt{p_0
q_0}}(\varepsilon^2(\textbf{q},\sigma)q^{1}-
\varepsilon^1(\textbf{q},\sigma)q^{2})F_m(\textbf{k}),\label{rsd401}
 \eea where
\bea F_m(\textbf{k})= \int_{V} e^{i{\bf k r}}B(\mathbf{r})
d\mathbf{r}. \eea In the cylindrical coordinates, the integral
(\ref{rsd401}) becomes \bea F_m(\textbf{k})= B\int^{R}_{0}\varrho
d\varrho\int^{2\pi}_{0}\exp \{i [k_x \cos\varphi +
k_y\sin\varphi]\} d\varphi\int^{l/2}_{-l/2}\exp \{ i k_z
z\}dz.\label{rsd4011} \eea After some manipulations we get
 \bea
F_m(\textbf{k})= \frac{4\pi
BR}{k_z\sqrt{k_x^2+k_y^2}}j_1\left(R\sqrt{k_x^2+k_y^2}\right)\sin\left(\frac{lk_z}{2}\right),\label{rsd41}
\eea  where $j_1$ is the spherical Bessel function of the first
kind.

 From Eqs.(\ref{rsd401},\ref{rsd41})  we obtain
the differential cross-section as follows \bea
\frac{d\sigma^m(\gamma\rightarrow\phi)}{d\Omega'}=\frac{2c^2_{\phi
\ga \ga}R^2 B^2 }{k_z^2(k_x^2+k_y^2)}
j_1^2\left(R\sqrt{k_x^2+k_y^2}\right)\sin^2\left(\frac{lk_z}{2}\right)(q_x-q_y)^2
\label{rsd42}. \eea Eq.(\ref{rsd42})  shows that when the momentum
of the photon is parallel to the z-axis (the direction of the
magnetic field), the differential cross-section vanishes. This
result is the same as the previous section. It implies that if the
momentum of the photon is parallel to the EM field, then there is
no conversion. If the momentum of the photon is parallel to the
x-axis, i.e., $q^\mu=(q,q,0,0),$ then  Eq.(\ref{rsd42}) gets the
form \bea \frac{d\sigma^m(\gamma\rightarrow\phi)}{d\Omega'}&=&
\frac{2c^2_{\phi \ga \ga}R^2 B^2 q^2
j_1^2\left(R\sqrt{(q-p\cos\theta)^2+(p\sin\theta
\cos\varphi')^2}\right)}{(p\sin\theta
\sin\varphi')^2[(q-p\cos\theta)^2+(p\sin\theta \cos\varphi')^2]}\nonumber\\
&&\times \sin^2\left(\frac{lp}{2}\sin\theta
\sin\varphi'\right),\label{rsd43} \eea where $\varphi'$ is the
angle between the y-axis and the projection of \textbf{p} on the
yz-plane [9].

Now we are interested in several cases. The first case is for
$\theta\approx 0$ we have \bea
\frac{d\sigma^m(\gamma\rightarrow\phi)}{d\Omega'}= \frac{c^2_{\phi
\ga\ga}R^2 B^2 l^2 j_1^2\left[Rq(1-
\sqrt{1-\frac{m^2_\phi}{q^2}})\right]}{2\left(1-\sqrt{1-\frac{m_\phi^2}{q^2}}\right)^2}
\label{rsd44}. \eea
 In the limit  $\theta\rightarrow\pi/2$ and $\varphi'\rightarrow 0$,  Eq.(\ref{rsd43}) becomes
\bea \frac{d\sigma^m(\gamma\rightarrow\phi)}{d\Omega'}=
\frac{c^2_{\phi \ga \ga}R^2l^2B^2 q^2}{2(2q^2-
m_\phi^2)}j_1^2\left(R\sqrt {2q^2-
m_\phi^2}\right).\label{rsd45}\eea For  the last  case, the limit
$ \theta\rightarrow\pi/2,\varphi'\rightarrow\pi/2$ yields \bea
\frac{d\sigma^m(\gamma\rightarrow\phi)}{d\Omega'}=
\frac{2c^2_{\phi \ga
\ga}R^2B^2}{(q^2-m_\phi^2)}j_1^2(Rq)\sin^2\left(\frac{lq\sqrt{1-\frac{m_\phi^2}{q^2}}}{2}\right).
\label{rsd46} \eea

\begin{figure}[h]\bc
\includegraphics[width=12cm,height=7cm]{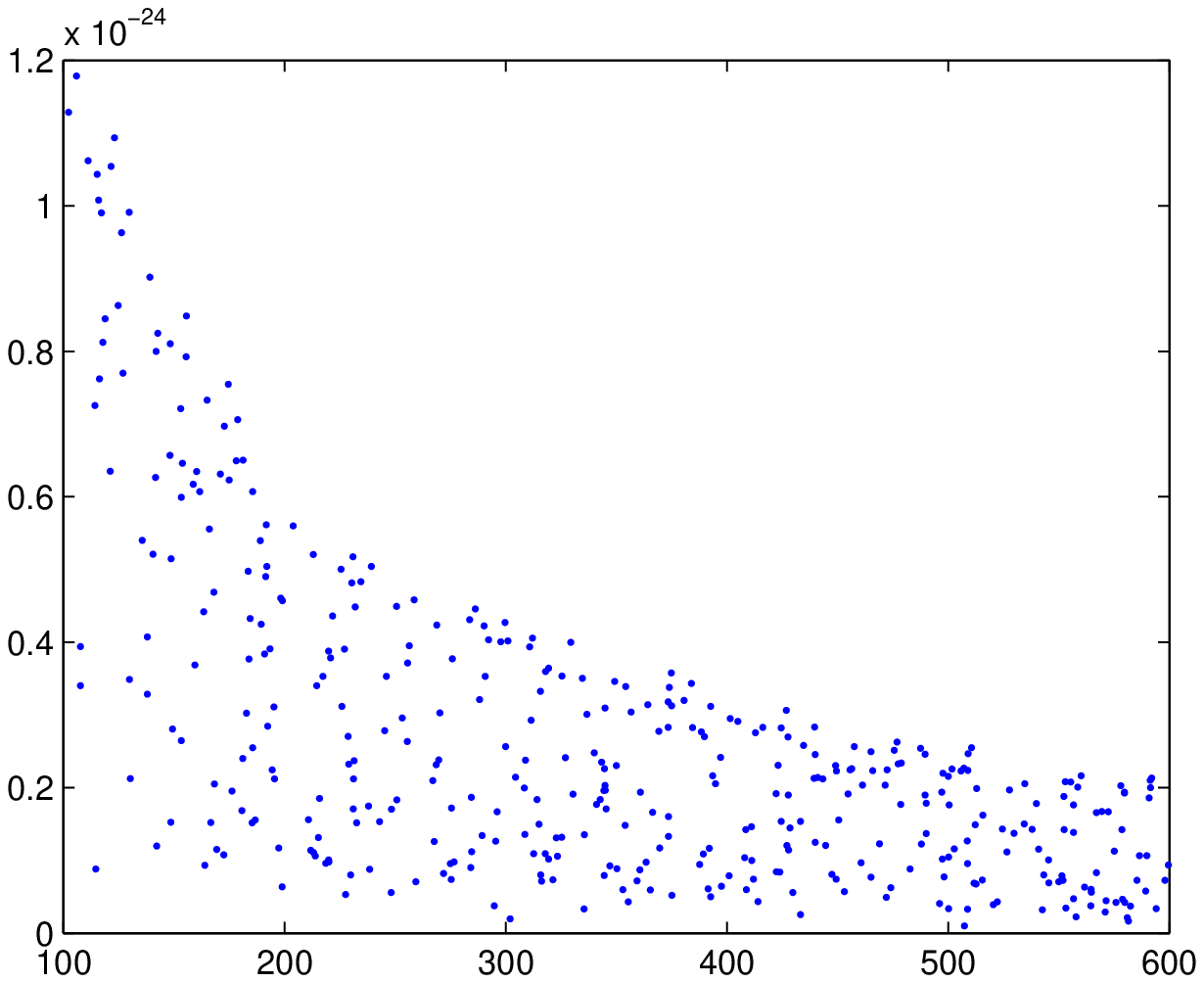}
\includegraphics[width=12cm,height=7cm]{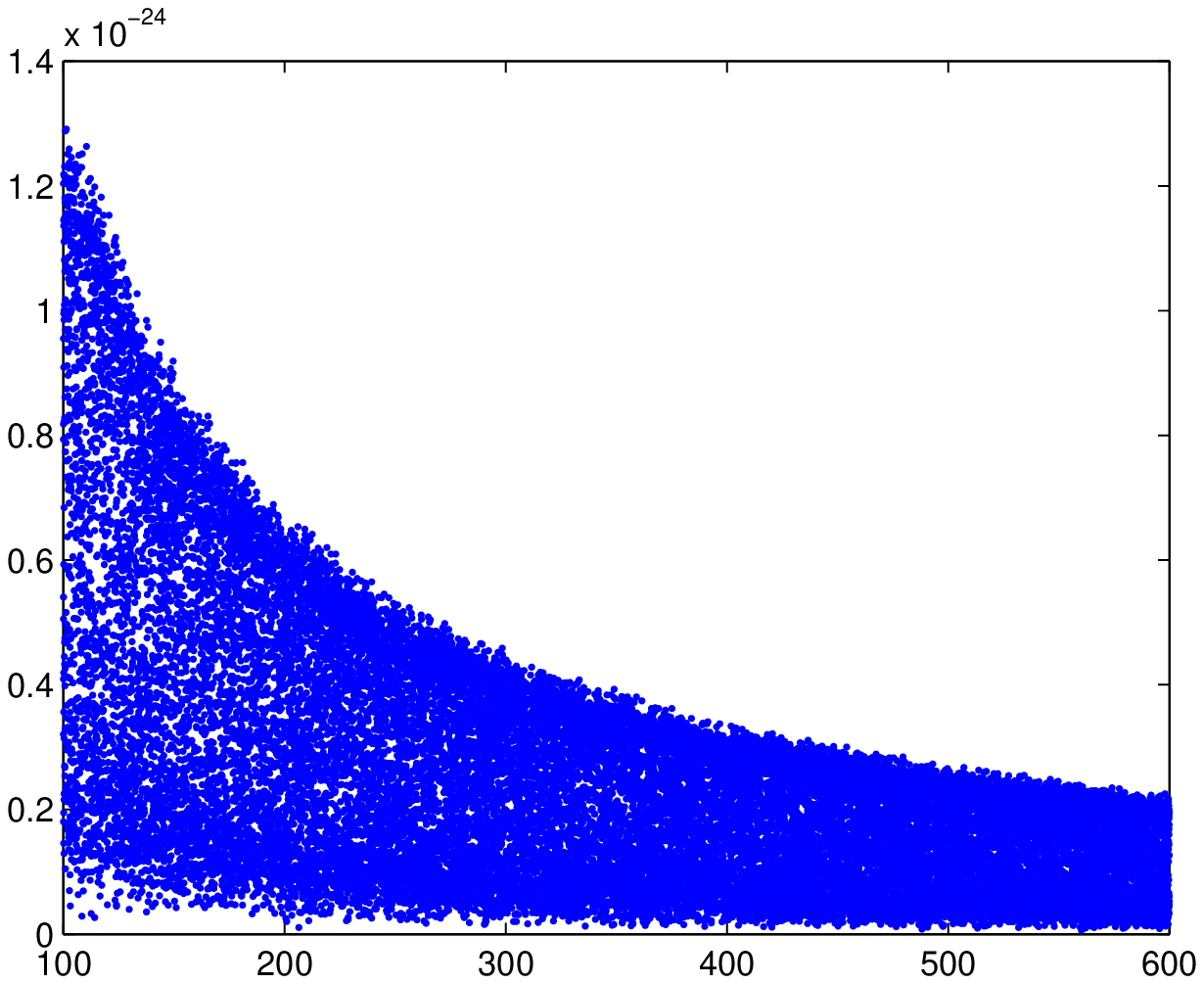}
\caption{\label{tab2} The total cross-section (cm$^2$) for
conversion in magnetic field as a function of provided momentum
$q=100-600$ GeV. The upper plot is depicted as 400 points, and the
lower one is for 30000 points.} \ec
\end{figure}

To evaluate the total cross-section from the general
formula (\ref{rsd43}), the parameter values for $\La_\phi$, $\al$
and $m_\phi$ are given as before. The remaining ones are chosen as
follows: $R=l=1\ \mathrm{m}=5.07\times 10^6\ \mathrm{eV}^{-1}$ and
$B = 9\ \mathrm{Tesla}=9\times 195.35\ \mathrm{eV}^2$ \cite{cast}.
The total cross-section on the selected range of momenta $q$ by
Eq.(\ref{rsd43}) for three cases $\xi =0, \pm \fr 1 6$ yields the
same value which is presented as in Figure \ref{tab2}.

From Fig. \ref{tab2} we see that the total cross-section for the
radion production in the  magnetic field are much bigger than that
of the electric field. This happens because of $B\gg E$. In this
case, the resonant figure disappears although the number of points
is much increased. This may be due to the fact that the
cross-section is given as an incoherent multiple of the Bessel and
trigonometric functions. Notice also that in similarity to the
previous case, the behavior of the cross-section is different in
the two domains divided by a transition point around on $200-300$
GeV. It is worth mentioning here if the radion mass is much
smaller than the provided photon momentum, the cross-sections are
much larger.

\subsection{Conversion in wave guide}

We will show that the cross-section can also be enhanced in
comparison to the electric field case when the conversion exists
in a varying EM field. Let us consider a case of the periodic
external EM field of the wave guide with the
$\mathrm{TE}_{10}$ mode. The nontrivial solution of this  mode is given by~\cite{jac} \\
\bea
H_z &=& H_o\cos\left(\frac{\pi x}{l_x }\right)e^{ikz-i\omega t},\nonumber \\
H_x &=& -\frac{ik l_x }{\pi}H_o\sin\left(\frac{\pi x}{l_x }\right)
e^{ikz-i\omega t},\nonumber \\
E_y &=& i\frac{\omega  l_x }{\pi}H_o\sin\left(\frac{\pi x}{l_x }\right)
e^{ikz-i\omega t}, \label{rsd45}
 \eea
 with  the cutoff frequency $\omega_o = \frac{\pi}{l_x }$.

The expression for the matrix element is \bea \langle p| {\cal M}
|q\rangle &=& \frac{c_{\phi\gamma\gamma}}{(2\pi)^{2}\sqrt{p_0
q_0}} \left\{ \varepsilon_2(\mathbf{q}, \tau)q_0 F_y+ \left[
\varepsilon_2(\mathbf{q}, \tau)q_3 - \varepsilon_3(\mathbf{q},
\tau)q_2\right]F_x\right.\crn &&\left.+\left[
\varepsilon_1(\mathbf{q}, \tau)q_2 - \varepsilon_2(\mathbf{q},
\tau)q_1\right]F_z \right\}, \label{rsd46}
 \eea
where $ p_0\equiv q_0 + \omega$,  and
\bea
F_x &=& - \frac{8 k l_x  H_0(q_x-p_x)\cos[\frac{1}{2}l_x (q_x-p_x)]
\sin[\frac{1}{2}l_y(q_y-p_y)]\sin[\frac{1}{2}l_z(q_z-p_z+k)]}
{\pi[(q_x-p_x)^2-\frac{\pi^2}{l_x ^2}](q_y-p_y)(q_z-p_z+k)},
\nonumber \\
F_y &=& - F_x,\label{rsd47} \\ F_z &=& -\frac{8\pi
H_0\cos[\frac{1}{2}l_x (q_x-p_x)]
\sin[\frac{1}{2}l_y(q_y-p_y)]\sin[\frac{1}{2}l_z(q_z-p_z+k)]} {l_x
[(q_x-p_x)^2-\frac{\pi^2}{l_x ^2}](q_y-p_y)(q_z-p_z+k)}.\nn
 \eea
 Substituting Eq.(\ref{rsd47}) into Eq.(\ref{rsd46}) we obtain the
differential cross-section
 \bea
\frac{d\sigma(\gamma \rightarrow \phi)}
{d\Omega}&=&\frac{c^2_{\phi\gamma\gamma}p_o}{2(2\pi)^2 q_0}\left[({q_0}^2 -
{q_y}^2){F_y}^2 + 2 q_0[(q_z - q_y)F_x+(q_y -{q_x}) F_z]F_y
+ (q_y - q_z)^2 {F_x}^2 \right. \nonumber \\
& &\left. +2[(q_z - q_y)q_y +(q_y -{q_z}) q_x ] F_xF_z
 +(q_x - q_y)^2 {F_z}^2\right].\label{rsd48}
 \eea

When the momentum of the photon is parallel to the z - axis, the
differential cross-section  vanishes. This is the same as in the
static EM fields. If the momentum of the photon is parallel to the
x - axis, Eq.(\ref{rsd48}) becomes \bea \fr{d\si(\ga \rightarrow
\phi )}{d \Omega} & = & \fr{8 c_{\phi \ga \ga}^2 H^2_0 l_x ^2q^2
}{ \pi^4} \left[\cos \fr{l_x (q-p \cos \theta)}{2}
 \sin \fr{l_y p \sin \theta \cos \varphi}{2} \sin \fr{l_z (p \sin \theta \sin
 \varphi+k)}{2}\right]^2\crn
&\times&\left(1 + \frac{\omega}{q}\right)\left[\omega(q-p \cos
\theta)+\frac{\pi^{2}}{l_x ^2}\right]^2 \left[(q-p \cos \theta)^2
-\frac{\pi^{2}}{l_x ^2} \right]^{-2}\crn &\times&\left[\left(
p\sin \theta \cos \varphi \right) (p\sin\theta\sin\varphi
+k)\right]^{-2}, \label{rsd49} \eea where $p=\sqrt{(q+\omega)^2
-m^2_{\phi}}$.

\begin{figure}[h]\bc
\includegraphics[width=12cm,height=7cm]{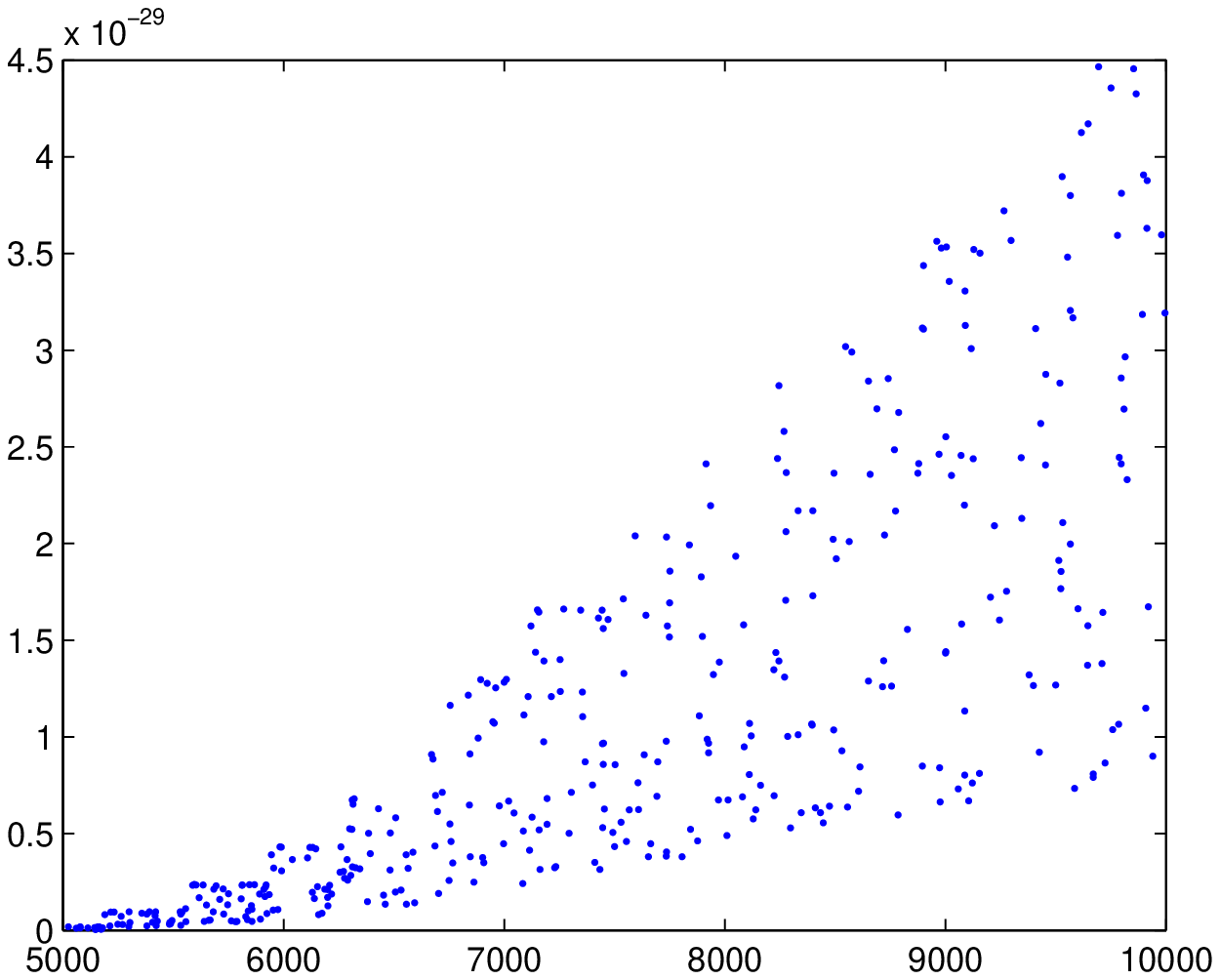}
\includegraphics[width=12cm,height=7cm]{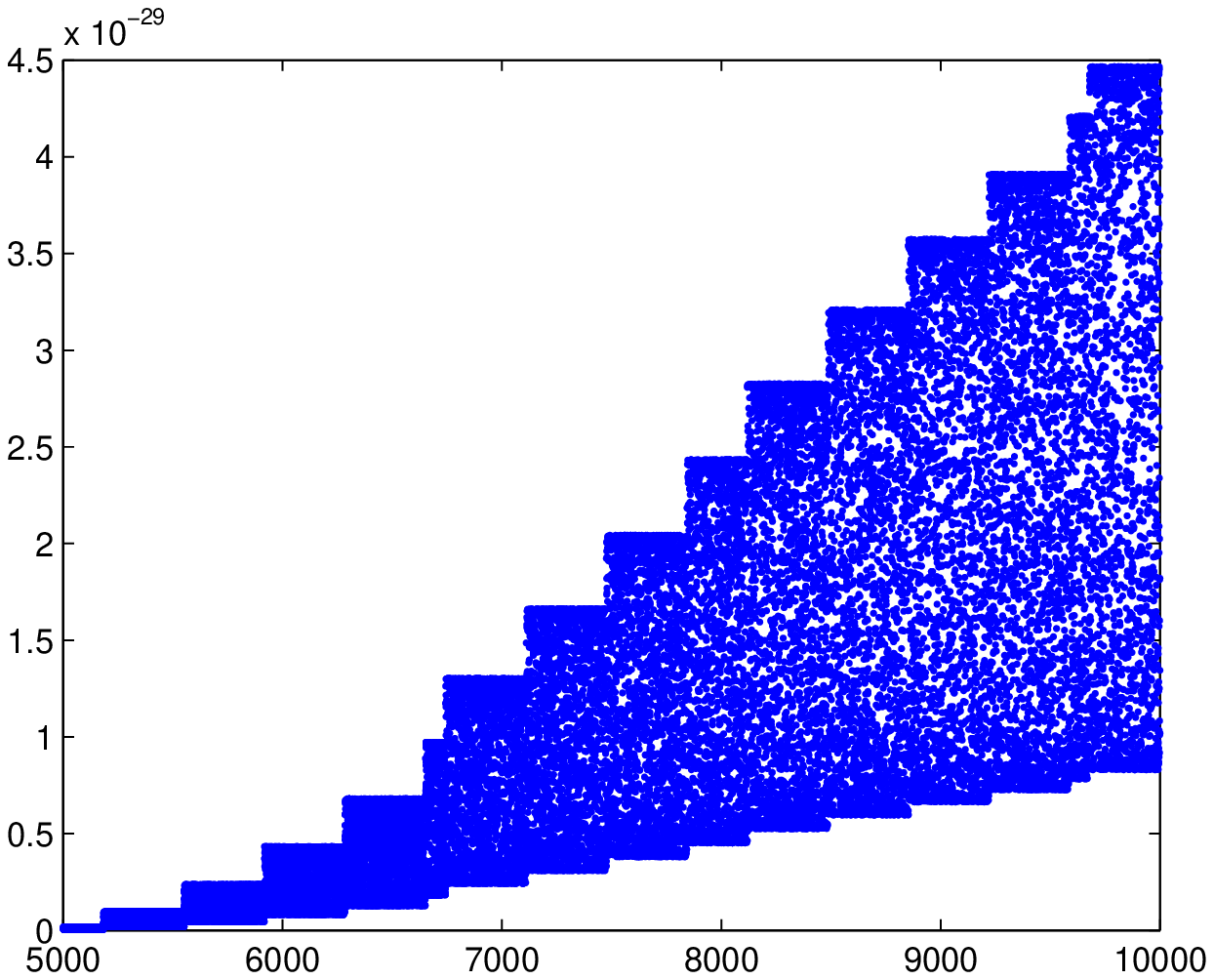}
\caption{\label{hinhvesua}The total cross-section (cm$^2$) as a
function of the external field frequency $
\omega/\omega_0=5000-10000$ corresponding to the value $q= 500\
\mathrm{GeV}$. The upper plot is depicted as 400 points, and the
lower one is for 30000 points.}\ec
\end{figure}

To compare the cross-section with the previous cases, we take $H_0
= B$, $q= 500 \ \mathrm{GeV}$ and $m_\phi=10\ \mathrm{GeV}$, for
example, into account. The remaining parameters are chosen as
before. In Fig.~\ref{hinhvesua}
we plot  the total cross-section as a function of the high
frequency external source: $\omega/\omega_0 = 5000 - 10000$. The
cross-section as presented is on average much larger than that of
the static electric field, but smaller than that of the static
magnetic field. It can become comparable to the latter when the
frequency is large enough. When the number of points plotted is
increased, the shape of steps become obvious. But, what do the
steps mean?

Let us first note that in this case the total cross-section
increases because it is proportional to $\om^2$ as can be
evaluated from (\ref{rsd49}) due to $\om_0 \ll \om,k \ll m_\phi
\ll p,q$. Since the provided photon source is fixed as mentioned,
any enhancement in the external EM field is correspondingly used
to convert into possible additionally-produced radions, which
depends drastically on the increasing rate and amount of the
external field energy momentum ($\om, \mathbf{k}$). The scale for
this energy momentum is proportional to
$k=|\mathbf{k}|\simeq\om\sim 10^3\om_0 \sim 10^{-3}$~eV, which is
so small in comparison to the radion mass. To create such a heavy
radion, $m_\phi = 10\ \mathrm{GeV}\gg 10^{-3}\ \mathrm{eV}$, an
appropriate change in (namely, a large enough amount of) $\om$
could be needed while the cross-section is still remained
constant. And, when the $\om$ approaches the upper bound of this
amount, the radion is just generated and the cross-section is then
stepped as seen in the plot. This happens similarly for the next
levels. Notice also that, in principle, this case could be
observed experimentally since one can control the frequency $\om$.

Let us remind that the cutoff frequency of the $\mathrm{TE}_{10}$
is $\omega_0 = \frac{\pi}{l_x }$ and at any given frequency $\om$
only a finite number of modes can propagate~\cite{jac}. It is
often convenient to choose the dimensions of the guide such that
in the operating frequency, only the lowest mode can occur. This
is an important point in order to apply it in experiments.

\section{\label{culc}Conclusion}
With the help of the coupling of radion to photons, we have
obtained the conversion cross-sections of photon into radion in
the presence of several external fields such as the static
electric field of the condenser, the static magnetic field of the
solenoid and the periodic electromagnetic field of the wave guide.
The numerical evaluations of the total cross-sections are also
given.

The production cross-sections of radion in the static electric
field is quite small, which is not expected to be easily observed.
However, the cross-section in the static magnetic and periodic
electromagnetic fields are much enhanced, which can be measurable
in the present experiments.

Let us mention again that since the Randall-Sundrum model radion
is quite heavy with masses at least in the $\mathrm{GeV}$ range,
the experiments are only available if the provided photon sources
are in high energies, as we often take some hundreds of GeV. Also,
the light radions in the model if they really exist are favored in
these experiments.

The original experiments were designed for searching axions and
gravitons which are the very light particles. Such possible light
radions if exiting in the model would be more accessible. In this
case, one can use our general formulae of the cross-sections, then
by the same procedure one can achieve the results, which have not
yet displayed in this work.

In this work we have considered only a theoretical basis for the
experiments, other techniques concerning construction and particle
detection can be found in Ref. \cite{cast}. It is emphasized that
our study can be applied for searching the possible light radions
in other models such as the large extradimensions.

 \section*{Acknowledgments}

We would like to thank the referees for their critical
remarks/comments which amended a lot the article. The work was
supported in part by National Foundation for Science and
Technology Development (NAFOSTED) of Vietnam.
\\[0.3cm]

\end{document}